\documentclass[a4paper]{PoS}

\makeatletter
\newcommand*{\rom}[1]{\expandafter\@slowromancap\romannumeral #1@}
\makeatother
\newcommand{\NNLOJET}{NNLO\protect\scalebox{0.8}{JET}}

\title{NNLO predictions for dijet production in diffractive DIS}

\ShortTitle{NNLO predictions for dijet production in diffractive DIS}

\author{\speaker{Radek \v{Z}leb\v{c}\'{i}k}\\
        DESY\\
        E-mail: \email{radek.zlebcik@desy.de}}

\author{Daniel Britzger\\
        Physikalisches Institut, Universi{\"a}t Heidelberg \\ 
        E-mail: \email{daniel.britzger@desy.de}}

\author{Jan Niehues\\
        Department of Physics, University of Z\"{u}rich\\
        E-mail: \email{jan.niehues@uzh.ch}}

\author{Thomas Gehrmann\\
        Department of Physics, University of Z\"{u}rich \\
        E-mail: \email{thomas.gehrmann@uzh.ch}}

\author{James Currie \\
        Institute for Particle Physics Phenomenology, Department of Physics, University of Durham \\
        E-mail: \email{james.currie@durham.ac.uk}}

\author{Alexander Huss \\
    Institute for Theoretical Physics, ETH, Z\"{u}rich \\
     E-mail: \email{ahuss@phys.ethz.ch}}

\abstract{
Cross sections for inclusive dijet production in diffractive deep-inelastic scattering are calculated for the first time in next-to-next-to-leading order (NNLO) accuracy.
These cross sections are compared to several HERA measurements published by the H1 and ZEUS collaborations.  
We computed the total cross sections, 49 single-differential and five double-differential distributions for six HERA measurements.
The NNLO corrections are found to be large and positive.
The normalization of the resulting predictions typically exceeds the data, while the kinematical shape of the data is described better at NNLO than at next-to-leading order (NLO).
Our results use the currently available NLO diffractive parton distributions, and the discrepancy in normalization highlights the need for a consistent determination of these distributions at NNLO accuracy.}

\FullConference{XXV International Workshop on Deep-Inelastic Scattering and Related Subjects\\
		3-7 April 2017\\
		University of Birmingham, UK}

\begin{document}

\section{Introduction}
Diffractive processes, $ep \to eXY$, where the systems $X$ and $Y$ are separated in rapidity, have been studied extensively at the electron-proton collider HERA~\cite{Aaron:2011mp,Andreev:2015cwa,Andreev:2014yra,Aktas:2007bv,Aktas:2007hn,Chekanov:2007aa}.
The forward system $Y$ usually consists of the leading proton but can also contain its low mass dissociation.
Between the systems $X$ and $Y$ is a depleted region without any hadronic activity, the so-called large rapidity gap (LRG), which is a consequence of the vacuum quantum numbers of the diffractive exchange, often referred to as a pomeron ($I\!P$).
Experimentally, the diffractive events can be selected either by requiring a rapidity region without any hadronic activity (LRG method) or by direct detection of the leading proton.
In the second case, the system $Y$ is free of any diffractive dissociation.

In analogy to the non-diffractive case, also in diffraction the parton distribution functions (PDFs) can be defined.
According to the factorisation theorem \cite{Collins:1997sr} the diffractive cross section is then expressed as a convolution of these diffractive PDFs (DPDFs) and partonic cross sections of the hard subprocess which are calculable within perturbative QCD (pQCD).
The DPDFs have properties similar to the PDFs, especially they obey the DGLAP evolution equation, but have an additional constraint on the presence of the leading proton in the final state.

The DPDFs are determined predominantly from inclusive diffractive deep-inelastic scattering (DDIS) cross sections with values of photon virtuality $Q^2$ much higher than $\lambda_{QCD}^2$, such that the factorisation theorem is valid.
In this way also the most commonly used DPDFs, H1 2006 Fit B \cite{Aktas:2006hy} were extracted.
The DPDFs are applicable to predict cross sections of other, more exclusive processes.
The most prominent one is inclusive dijet production in DDIS, where at least two jets are produced, mainly via boson-gluon fusion mechanism.
Since the gluonic component of DPDFs is weakly constrained from inclusive DDIS data alone also dijet data are sometimes considered in their determinations~\cite{Aktas:2007bv,Chekanov:2009aa}.

The pQCD predictions depend on the unphysical renormalisation and factorisation scales.
In inclusive DIS these scales are typically identified with the photon virtuality $Q^2$ and the cross section scale uncertainties are quite small.
On the contrary, the dijet production is a multi-scale process, where the presence of another hard scale, given for instance by the jet transverse momenta, makes a sensible scale choice much more complex.
The scale uncertainties of the cross section are typically much larger and represent the main theoretical uncertainty.
Therefore, the higher-order perturbative predictions, which reduce scale-dependence, are of the vital importance. 

Up to now, the measured dijet data were typically compared to the next-to-leading (NLO) QCD predictions which were within large theoretical uncertainties able to describe the measured cross sections satisfactorily, both in shape and normalization.
However, since these predictions were about two times higher than the leading-order (LO) one, there was a natural question concerning the size of contributions from even higher orders.
To address this issue, we present the next-to-next-to-leading (NNLO) perturbative QCD calculations.
These calculations are performed for the first time. 
We compare our predictions with several single-, double-differential and total cross sections from six~distinct measurements published by the H1 or ZEUS.
We further study the scale dependence of the NNLO predictions and investigate different DPDFs parametrisations.

\section{Variable definition}
\begin{figure}[h]
\centering
\includegraphics[width=0.4\textwidth]{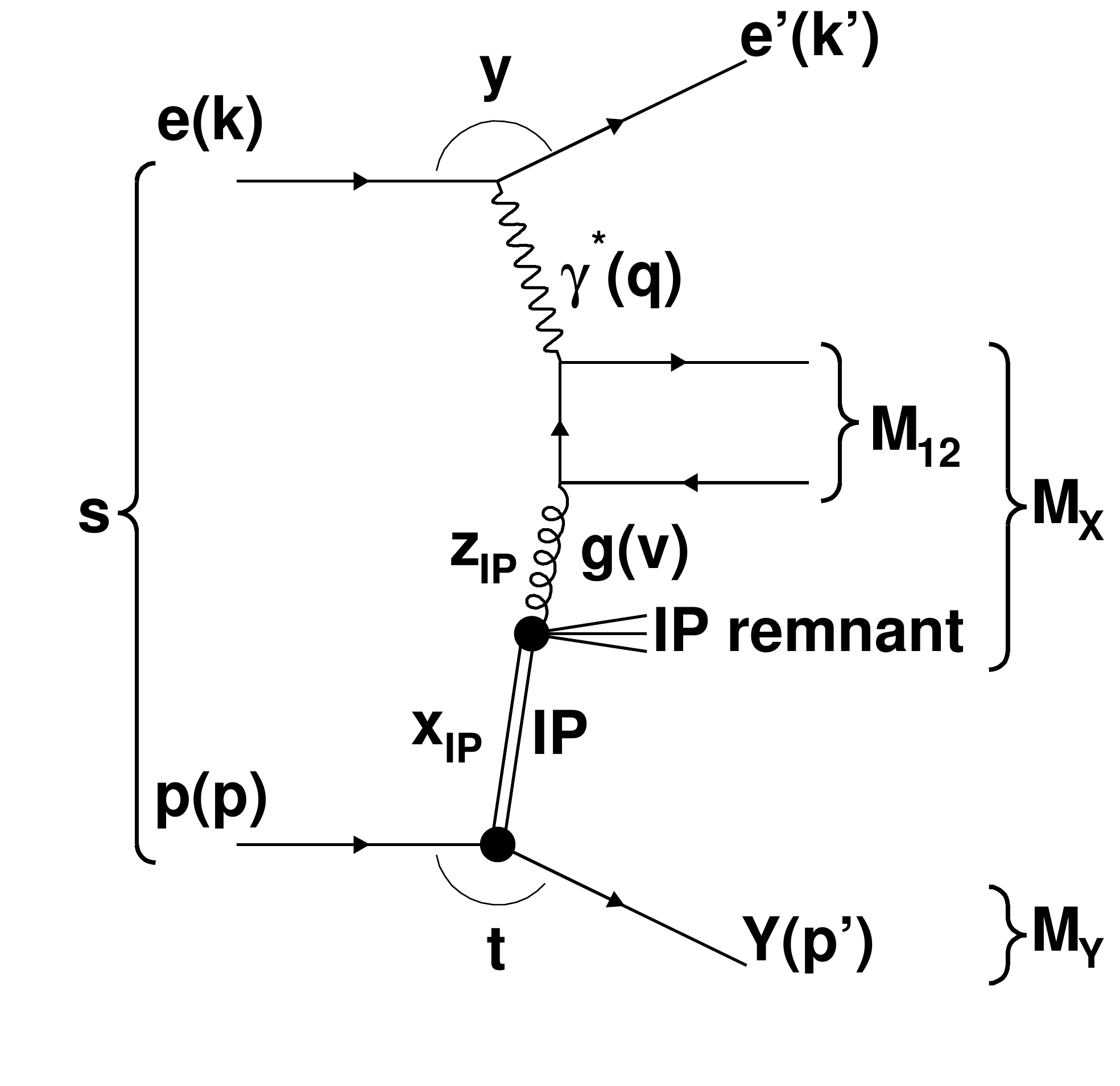}
\caption{The leading order Feynman diagram of the dijet production in diffractive DIS (taken from \cite{Andreev:2014yra}).}
\label{diagFeyn}
\end{figure}

Dijet cross sections are studied differentially in several kinematic variables, which also constrain the phase space of the measurements.
Their meaning is described in Fig.~\ref{diagFeyn}.
The jets were always identified using the $k_T$-algorithm in the $\gamma^*p$ frame with the parameter $R=1$.
The collaborations measured jet properties, like transverse momenta and pseudo-rapidities of the jets, either in $\gamma^*p$ or in the laboratory frame.
In almost all analyses the diffractive variables $x_{I\!P}$ and $z_{I\!P}$, where $x_{I\!P}$ is the relative energy loss of the beam proton caused by the diffractive scattering, and $z_{I\!P}$ which is interpreted as the momentum fraction of the parton entering the hard subprocess with respect to the diffractive exchange (pomeron), were measured.
The exact definitions of these variables can be found in \cite{Andreev:2014yra}.

\section{NNLO calculations}
Our theoretical NNLO QCD predictions for dijet production in DDIS are calculated using the program \NNLOJET{} and are based on antenna subtraction method~\cite{Currie:2016ytq,Currie:2017tpe}.
In this way the infrared divergences from real-real, real-virtual and virtual-virtual contribution are correctly handled using local subtraction terms.
The NNLO calculations were recently successfully used for predictions of jet cross sections in non-diffractive DIS \cite{Andreev:2016tgi}.
The NLO 2jet and 3jet cross sections were verified against Sherpa \cite{Gleisberg:2008ta} and NLOJET++ \cite{Nagy:2003tz}. 

In order to compare the data with fixed-order predictions, correction factors to account for hadronisation effects have to be applied and these are provided for 5 out of the 6 analysed measurements by the collaborations.
These factors are not provided to us in case of ref.~\cite{Chekanov:2007aa}  and we use the unity factor instead.

As the NNLO calculations are very computation-time consuming (more than 100,000 CPU hours) we are using the fastNLO framework \cite{Britzger:2012bs} to perform the final convolution of the hard subprocess cross section with the DPDFs and $\alpha_{s}$.
The advantage of this approach is that the computationally intense calculation of the hard-process cross sections for values of $x$, $Q^2$ and $p_T$ is performed only once.
Subsequent to that the partonic cross sections can be easily convoluted with different DPDFs and $\alpha_s$ values.
This approach is especially beneficial for performing PDF fits or fits of the strong coupling constant.
Note, that for historical reasons the program is called fastNLO although the idea behind is rather general and can be applied to all orders. 

For the diffractive processes, the convolution of DPDFs and the partonic cross sections is more complicated as in case of non-diffractive DIS, since the DPDFs depends on $x$, $\mu$ and $x_{I\!P}$\footnote{We integrated out an additional dependence on $t$}.
Therefore, for a given $x$, one needs to perform an additional convolution over $x_{I\!P}$, in general from $x$ to 1, but due to the $x_{I\!P}$ range of the given measurement the resulting range of the integration is smaller (mostly $x_{I\!P} < 0.03$).
This method has been found to be equivalent to the commonly used slicing method \cite{Aktas:2007hn,Andreev:2014yra}.

\section{Results}
In total we computed the NNLO cross sections for the six HERA measurements.
In plots they are labeled as:
"H1 (HERA \rom{2}) FPS" \cite{Aaron:2011mp}, "H1 (HERA \rom{2}) VFPS" \cite{Andreev:2015cwa}, "H1 (HERA \rom{2}) LRG" \cite{Andreev:2014yra}, "H1 (HERA \rom{1}) LRG" \cite{Aktas:2007bv}, "H1 (HERA \rom{1}) LRG, 820GeV" \cite{Aktas:2007hn}, "ZEUS (HERA \rom{1}) LRG" \cite{Chekanov:2007aa}.
Five of them are performed with proton beam energy of 920~GeV, one of them with 820~GeV.
The electron beam energy is always 27.6~GeV.
In two cases the leading proton is identified by the Forward Proton Spectrometer (FPS)~\cite{Aaron:2011mp} or Very Forward Proton Spectrometer (VFPS)~\cite{Andreev:2015cwa}, otherwise the diffractive events are selected using the LRG method.
The phase space definitions are found in the corresponding publications.

\begin{figure}[h]
\centering
\begin{minipage}{.49\textwidth}
\includegraphics[width=1.0\textwidth]{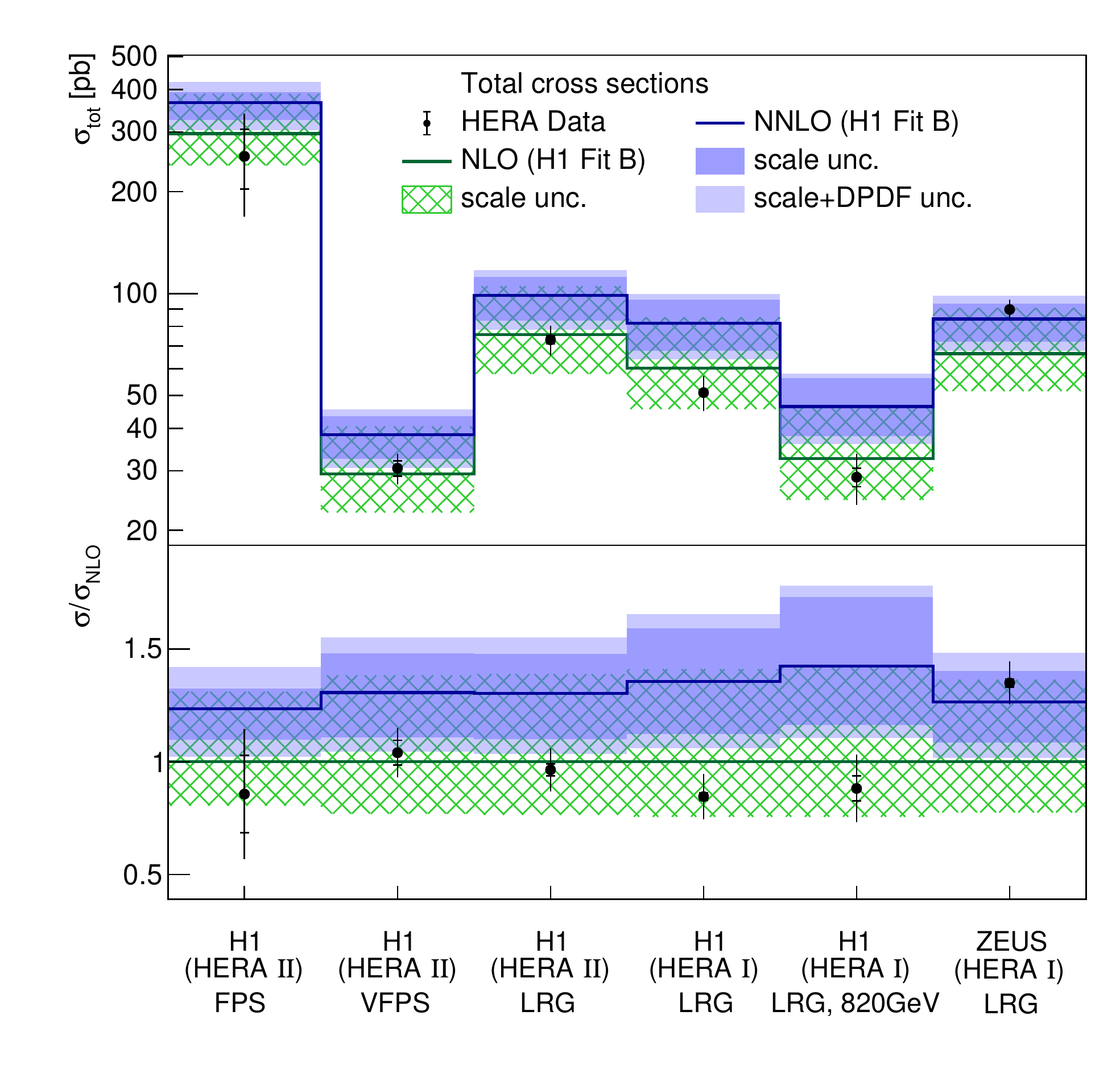}
\end{minipage}
\begin{minipage}{.49\textwidth}
\includegraphics[width=1.0\textwidth]{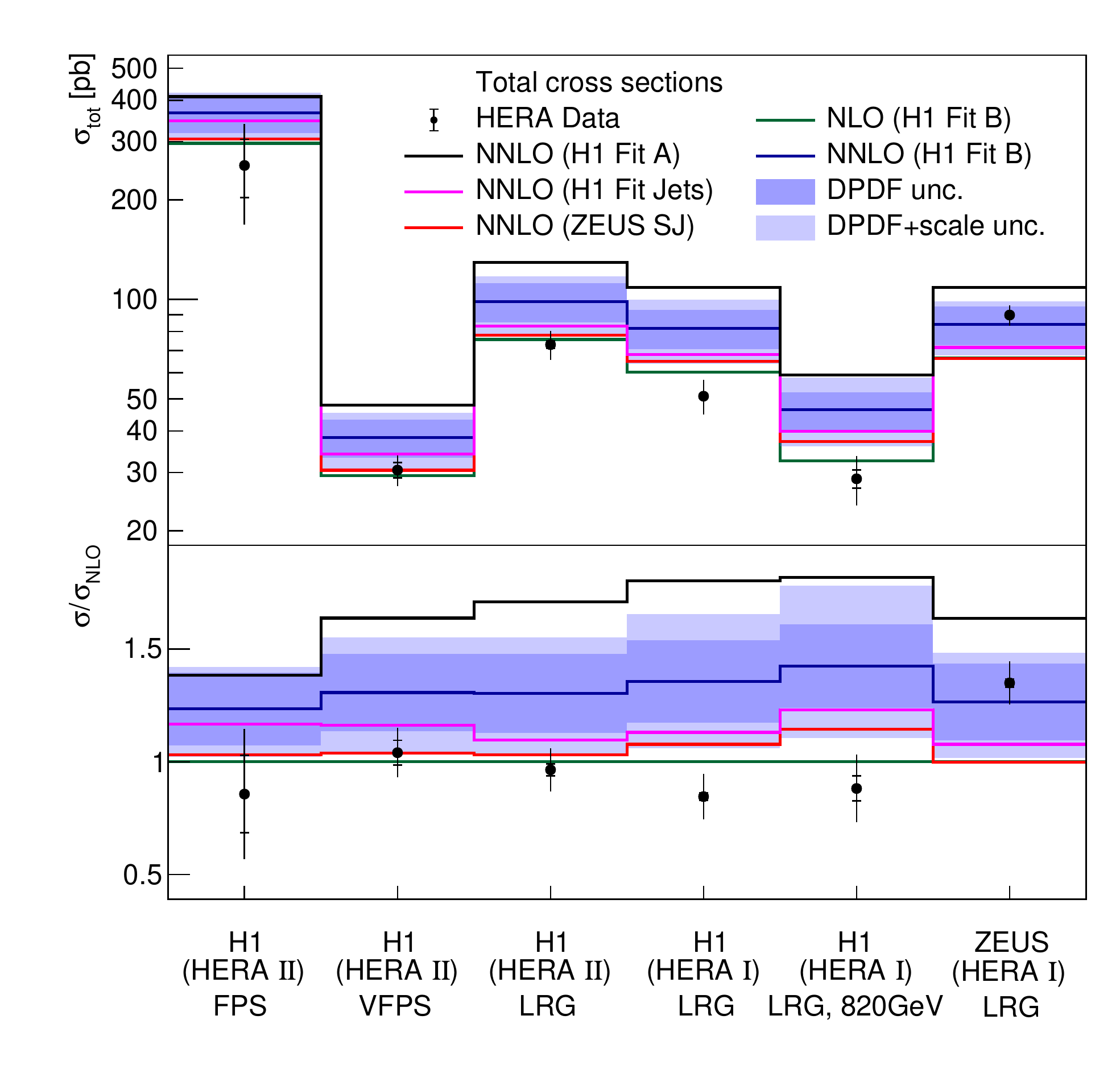}
\end{minipage}
\caption{The comparison of the total cross sections of all analysed measurements with theoretical QCD predictions at NLO and NNLO accuracy. The inner data error bars represent statistical uncertainties and other error bars are statistical and systematic errors added in quadrature. On the left, the theoretical predictions using H1 2006 DPDF Fit B are shown with the scale uncertainties (NLO and NNLO) and with scale and DPDF uncertainties added in quadrature (NNLO). The right plot compares NNLO predictions using several DPDF fits. For H1 2006 Fit B NNLO QCD predictions the DPDF and scale uncertainties are depicted.}
\label{figTotXsec}
\end{figure}

First, we present the results for total cross sections in the left plot of Fig.~\ref{figTotXsec} and compare to measurements and previously employed NLO calculations .
The NNLO predictions are higher than the NLO predictions and data (with the exception of ZEUS measurement) by about 30\%.
As found previously \cite{Aaron:2011mp,Andreev:2015cwa,Andreev:2014yra,Aktas:2007bv,Aktas:2007hn,Chekanov:2007aa}, the NLO predictions are mainly in good agreement with data.
The scale uncertainties, obtained by simultaneous variation of the renormalisation and factorisation scale by a factor of $0.5$ and $2$, are found to be somewhat reduced for NNLO predictions as compared to NLO.

In Fig.~\ref{figTotXsec} (right) we study the dependence of the total cross sections on the DPFS,
using H1 2006 Fit A~\cite{Aktas:2006hy}, H1 2006 Fit B~\cite{Aktas:2006hy}, H1 2007 Fit Jets \cite{Aktas:2007bv} and ZEUS SJ \cite{Chekanov:2009aa}.
The NNLO prediction mainly overshoot the data also for different DPDFs.
However, it is observed that the DPDFs considering also dijet data in their determination~\cite{Aktas:2007bv,Chekanov:2009aa} give smaller predictions than the inclusive-only fits.
The differences between the predictions are mostly covered by the DPDF uncertainties of the H1 2006 Fit B.
The DPDF H1~2006~Fit~A~\cite{Aktas:2006hy}, which considers only inclusive data in its determation, appears to overestimate the gluon component significantly.

\begin{figure}[h]
\centering
\begin{minipage}{.49\textwidth}
\includegraphics[width=1.0\textwidth]{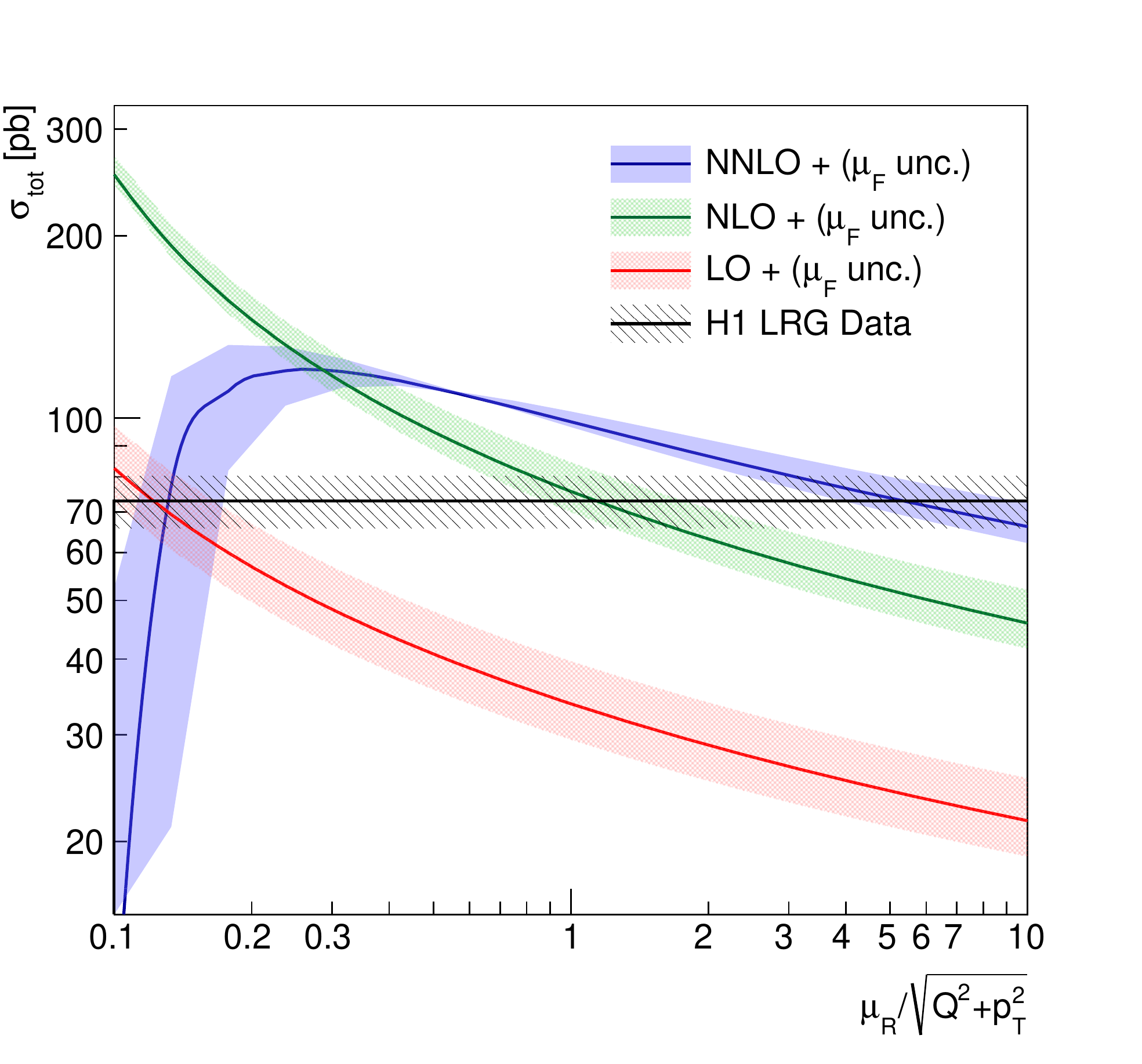}
\end{minipage}
\begin{minipage}{.49\textwidth}
\includegraphics[width=1.0\textwidth]{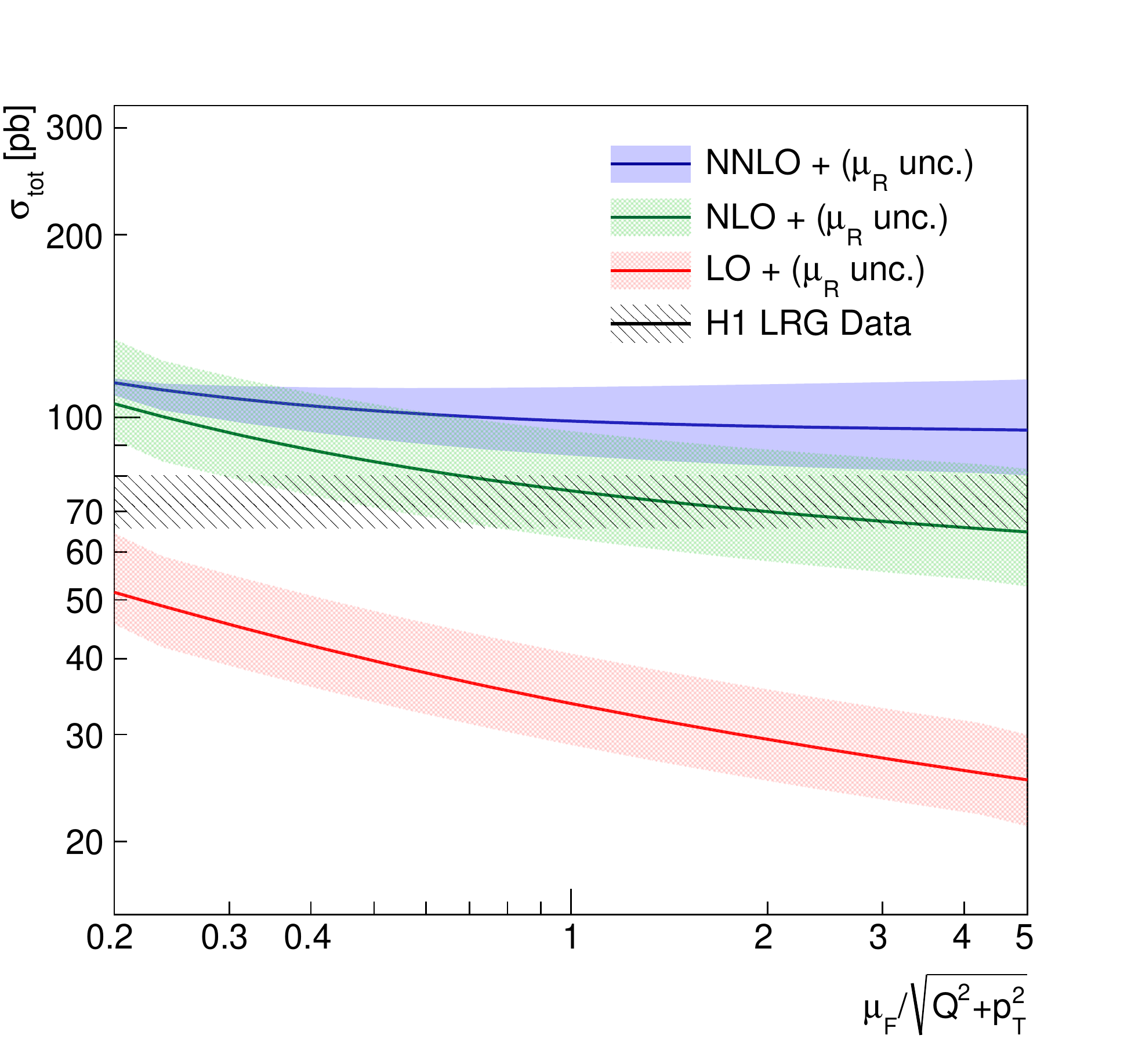}
\end{minipage}
\caption{The dependence of the total cross section of the H1 HERA \rom{2} LRG analysis \cite{Andreev:2014yra} on the renormalisation (left) and factorization (right) scale. 
The calculated cross sections are shown at LO, NLO and NNLO accuracy.
In the left (right) plot the uncertainties from factorisation (renormalisation) scale variation by the factor of 2 are depicted by the color band.
As a reference also the measured data cross section with its uncertainties is plotted.}
\label{figScaleDependence}
\end{figure}

We study the dependence of the total cross section on the renormalisation and factorisation scales $\mu_R$ and $\mu_F$ for the H1 LRG phase space in Fig.~\ref{figScaleDependence} (left).
The nominal value of the scales is $\mu_F^2 = \mu_R^2 = Q^2 + \langle p_T \rangle^2$, where $\langle p_T \rangle$ denotes the average transverse momentum of the leading and sub-leading jet. 
In the left plot we varied $\mu_R$ by factors between $0.1$ and $10$, while $\mu_F$ was fixed to its nominal value.
The effect of the variation of $\mu_F$ is displayed by an error band corresponding to a variation by factors of $0.5$ and $2$.
Fig.~\ref{figScaleDependence} (right) shows the same but with $\mu_R$ and $\mu_F$ swapped.
The NNLO cross sections are found to be less dependent on scale variations as compared to LO or NLO prediction.
The prediction exceeds the H1 HERA \rom{2} data for a wide range of the scale factors.

\begin{figure}[h]
\includegraphics[trim={1.9cm 0 0 0},clip,width=\textwidth]{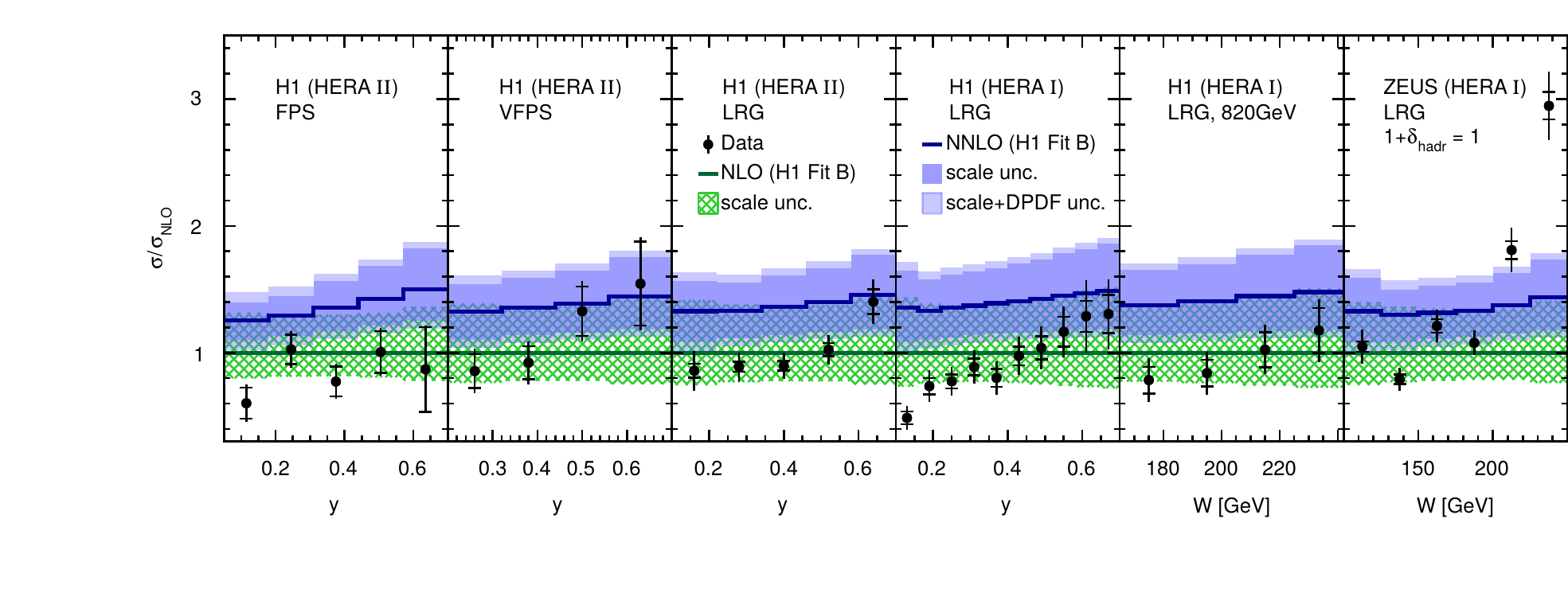}
\includegraphics[trim={1.9cm 0 0 0},clip,width=\textwidth]{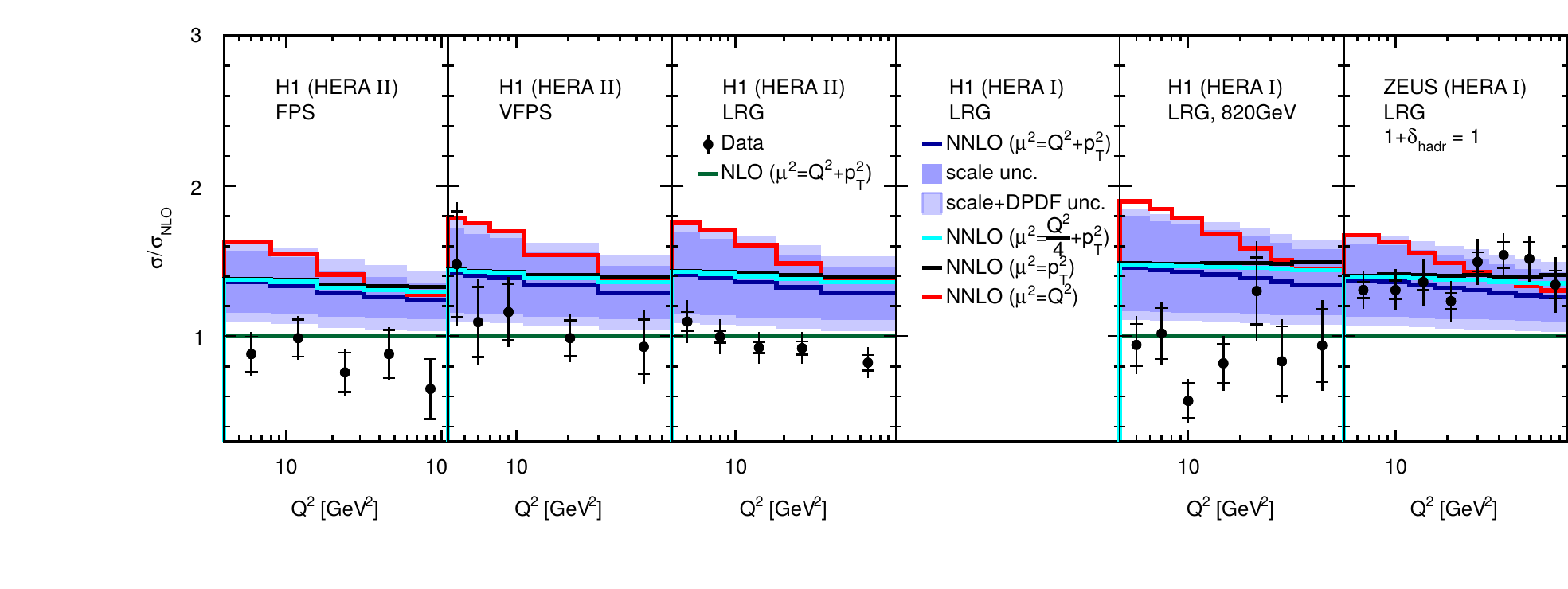}
\includegraphics[trim={1.9cm 0 0 0},clip,width=\textwidth]{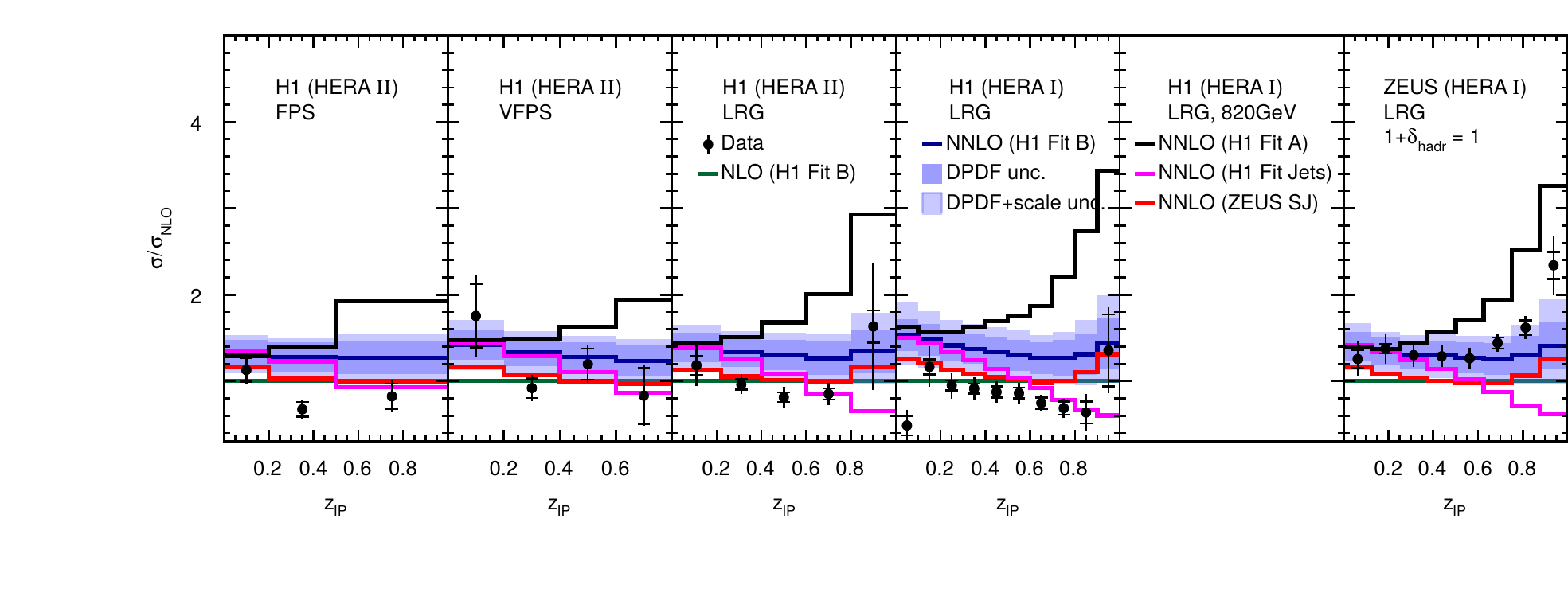}
\caption{The ratio of the differential cross sections with respect to the NLO predictions for $y$ and $Q^2$ and $z_{I\!P}$ variables. In some cases instead of $y$ the equivalent variable $W$ was measured.  The QCD predictions are corrected for the effect of hadronisation using published hadronisation correction. Only in case of ZEUS measurement where no hadronisation corrections were published we assume $1+\delta_{hadr}=1$. For $Q^2$ distribution several possible choices for the renormalisation and factorization scale are studied. For the $z_{I\!P}$ distribution several DPDFs were used in the NNLO QCD calculations.}
\label{figDiffDep}
\end{figure}

In total we computed 57 differential distributions.
Fig.~\ref{figDiffDep} shows the distributions for inelasticity $y$, $Q^2$ and $z_{I\!P}$ normalised to the respective NLO predictions.
For $y$, which is related to the $\gamma^*p$ centre-of-mass energy $W\simeq\sqrt{ys}$, the NNLO predictions provide an improved description of the shape as compared to NLO, while still being too high in normalisation.

Together with $Q^2$ distributions the predictions for various functional forms of the hard scale are displayed in addition. We use $\mu = \mu_F = \mu_R$.
The prescriptions containing mean $p_T$ of the dijets $\mu^2 = Q^2/4 +\langle p_T\rangle^2$, $\mu^2 = Q^2 + \langle p_T\rangle^2$ and $\mu^2 = \langle p_T\rangle^2$ give all similar results, whereas the scale choice $\mu^2 = Q^2$ results in higher cross sections and a steeper $Q^2$ spectrum.
However, all these scale choices are covered by the scale uncertainty.

The bottom pad of Fig.~\ref{figDiffDep} shows the $z_{I\!P}$ variable which represents in leading order approximation one of the DPDF arguments.
The NNLO predictions exceed the data for most of the $z_{I\!P}$ measurements.
The H1 2006 Fit A overestimates the cross section for higher $z_{I\!P}$ values significantly whereas the H1 2007 Fit Jet DPDFs gives the smallest cross section.
The remaining two DPDFs (H1 2006 Fit B and ZEUS SJ) behave quite similar.
There is an indication that the NNLO predictions describe data distributions better in shape, in particular, we found that for every DPDF and every studied scale functional form the NNLO predictions on average gives lower $\chi^2$ than NLO predictions.

\section{Conclusions}
We present the first NNLO QCD predictions for diffractive jet production processes. 
These were calculated for dijet production in diffractive DIS and were confronted to six measurements by H1 or ZEUS.
We observe that the NNLO cross sections are typically significantly higher than the data and are about $30\%$ higher than NLO calculations. 
Since no DPDFs in NNLO accuracy exist, only available NLO DPDFs were used for the calculations which may already explain the discrepancy between the NNLO predictions and data.
The NNLO predictions exceed the data also for all DPDFs studied.
However, the shapes of the differential distributions are typically better described by NNLO predictions, which was quantified by $\chi^2$-calculations.
The NNLO predictions provide a reasonable description of the ZEUS data, while, however, these are in general found to be higher than the five H1 measruementes and no hadronisation corrections are available to us.
We believe that the normalisation difference between data and NNLO predictions could be explained by the inconsistent order of the employed NLO DPDFs,
and new DPDFs in NNLO accuracy, including dijet data for their determinations, would be highly recommended.

\bibliographystyle{JHEP}
\bibliography{refs}

\end{document}